\documentclass[useAMS,usegraphicx]{mn2e}

\title[Orientation Effects in Quasars]{
[O~III] Equivalent Width and Orientation Effects in Quasars
}
\author[G. Risaliti et al.]
{G. Risaliti,$^{1,2}$ 
M.~Salvati,$^1$ \& A. Marconi$^3$ \\
$^1$ INAF - Osservatorio Astrofisico di Arcetri, L.go E. Fermi 5,
Firenze, Italy\\
$^2$ Harvard-Smithsonian Center for Astrophysics, 60 Garden St. 
Cambridge, MA 02138 USA {E-mail: grisaliti@cfa.harvard.edu}\\
$^3$ Dipartimento di Fisica e Astronomia, Universit\`a di Firenze, Largo E.~Fermi 2, Firenze, Italy\\
}

\begin{document}

\date{Released Xxxx Xxxxx XX}

\pagerange{\pageref{firstpage}--\pageref{lastpage}} \pubyear{2002}

\maketitle

\label{firstpage}

\begin{abstract}
The flux of the [OIII] $\lambda$5007\AA~ line is considered to be a good indicator
of the bolometric emission of quasars. The observed continuum emission from the
accretion disc should instead be strongly dependent on the inclination angle $\theta$
between the  disc axis and the line of sight.
Based on this, the equivalent width (EW) of $\rm [OIII]$ should provide a direct 
measure of $\theta$.
Here we analyze the distribution of EW([OIII]) in a sample of $\sim$6,000 SDSS
quasars, and find that it can be accurately reproduced assuming a relatively small intrinsic scatter
and a random distribution of inclination angles.
This result has several implications: 1) it is a direct proof of the disc-like
emission of the optical continuum of quasars; 2) the value of EW([OIII])
can be used as a proxy of the inclination, to correct the measured continuum emission and
then estimate the bolometric luminosity of quasars; 3) the presence of almost edge-on
discs among broad line quasars implies that the  accretion disc is not
aligned with the circumnuclear absorber, and/or that the covering fraction of the latter 
is rather small. Finally, we show that a similar analysis of EW distributions of 
broad lines (H$\beta$, Mg~II, C~IV) provides no evidence of inclination effects,
suggesting 
a disc-like geometry of
the broad emission line region.

\end{abstract}


\section{Introduction}

The primary optical/UV emission of quasars is thought to arise from an accretion disc  surrounding a supermassive black hole.
Radiatively efficient discs (as expected in quasars, Marconi et al.~2004, and
described by the Shakura \& Sunyaev~1972 model),
are geometrically thin and optically thick. Therefore, their observed emission
should scale with the cosine of the disc inclination angle with respect to the
line of sight. This is a basic element of any model of quasar emission, yet
it is rather difficult to test directly, given the uncertainties on the measurements
of the intrinsic continuum emission. 

The obvious way to perform a check of the disc geometry of the optical/UV
emitter is through a comparison with an inclination-independent indicator of the
intrinsic quasar luminosity.

Such an indicator should have:
1) a negligible contamination from processes other than AGN emission (such as
nuclear star formation), and
2) a small scatter in its functional dependence from the bolometric luminosity.

The first requirement is matched by several observables, for instance the hard X-ray flux,
the flux of broad emission lines, and that of high-ionization narrow emission lines,
such as [OIII]~$\lambda$~5007~\AA\ (Mulchaey et al.~1994), [OIV]~$\lambda$~24.5~$\mu$m
(Rigby et al.~2009).  
The second requirement is hard to quantify, due to the lack of independent ways
to measure the intrinsic bolometric emission. Therefore, the calibration of 
these indicators is not an absolute one, and is instead based on internal consistency.

In this paper we propose an easy and direct way to (a) calibrate the
[OIII] flux as an indicator of the intrinsic luminosity, and (b) to 
estimate the inclination effects on the continuum emission of quasars, 
based on the analysis of the distribution of the [OIII] equivalent width (EW)
in the Sloan Digital Sky Survey (SDSS) DR5 quasar sample (Schneider et al.~2007).

\begin{figure}
\includegraphics[width=8.5cm,angle=0]{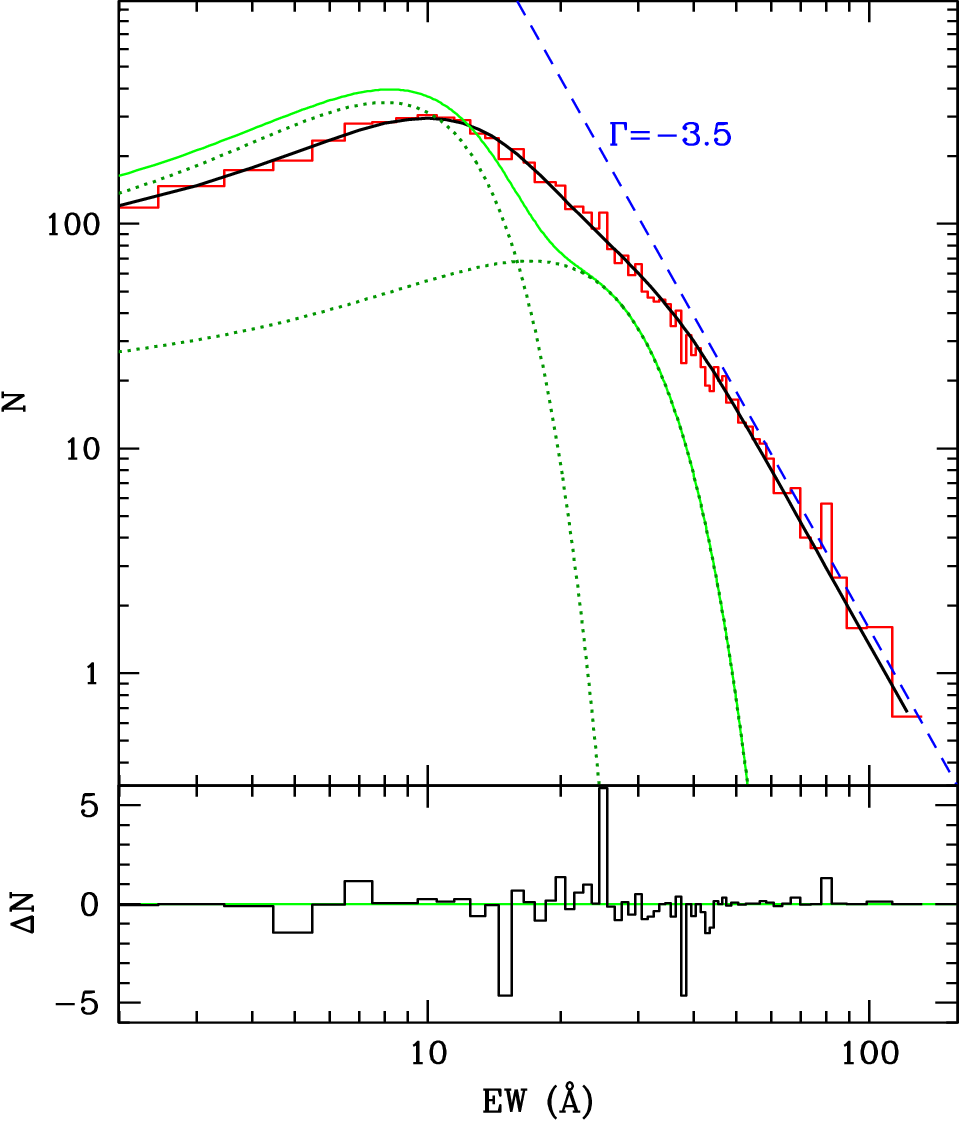}
\caption{Distribution of EW([OIII]) (red histogram) and best fit model (black continuous line)
for a sample of 6,029 SDSS quasars, as 
defined in the text. The dashed blue line is a power law with slope $\Gamma$=-3.5 and arbitrary normalization, 
shown for ease of comparison with the slope of the high-EW tail. The continuous light green curve is the
total intrinsic distribution, obtained adding two Gaussian components (the dark green, dotted curves). 
 The continuous light green and black curves have the same normalization.}
\label{ttfit1}
\end{figure}

\section{General method and sample selection}
The starting point of our analysis is the following, straightforward
consideration, valid for an "idealized" quasar sample: 
if the [OIII] luminosity is a "perfect" indicator of
the intrinsic luminosity, and the continuum emission is
due to an optically thick disc, then the distribution of
[OIII] equivalent widths (EW) observed in a non-biased sample of
quasars should simply reflect the distribution of disc orientations
(this further assumes a negligible spread in the quasar spectral
energy distribution (SED), so that the continuum luminosity at the line energy 
is also perfectly correlated with the total luminosity).
In particular, if $\theta$ is the angle between the disc axis and the
line of sight, we have EW$_{O}$=EW$^*/\cos\theta$, where EW$_{O}$ is the 
observed equivalent width, and EW$^*$
is the equivalent width as measured in
a face-on disc, which in this idealized case is a fixed value for all quasars.
For a population of randomly oriented discs, we expect the same
number of objects per element of solid angle,
and, being $\cos\theta$=EW$^*$/EW$_{O}$, the observed distribution
of [OIII] equivalent widths should be (Netzer~1985,~1990):
\begin{equation}
dN\propto d\Omega=d(\cos \theta)\, d\phi = \frac{EW^*}{EW_O^2}d(EW_O)\, d\phi
\end{equation}

Next we estimate
how the expected distribution changes if one moves from the idealized case
described above to the "real world". Two main points must be considered:
the possible intrinsic spread of the [OIII]--bolometric emission relation, and
the selection effects inherent to the given sample. \\

{\bf 1. Intrinsic spread of the [OIII]--bolometric luminosity relation.}
Our working hypothesis is that on average the [OIII] luminosity is indeed a good indicator of
the bolometric luminosity. This is supported by several observational studies, where
the [OIII] flux is compared with the optical and X-ray emission (e.g. Mulchaey et al.~1994, Heckman
et al.~2005),
and by studies of the SDSS sample, where it is found that the [OIII] line is dominated by the
AGN contribution, with no significant contribution by the host galaxy, even in cases
of strong star formation activity (Kauffmann et al.~2003).
However, the spread between the emission in [OIII]~ and in other bands is large
(Heckman et a.~2005). This is expected, since the flux in a narrow line depends
on several variable parameters, including 
the intensity and shape of the ionizing continuum, and the geometry, distance
and covering factor of the narrow line region (e.g., Baskin \& Laor~2005).
A realistic description of the expected distribution of EW([OIII]) must therefore allow
for some intrinsic spread. We modeled this distribution as a Gaussian, to be convolved with the
effect of inclination described above. The goodness of the fit to the observed
distribution, and the ratio between the distribution width and the fiducial value EW$_0$ will
assess the goodness of our hypothesis, and, in particular, of the accuracy of
the [OIII] luminosity as a proxy of the total quasar luminosity.\\
{\bf 2. Selection effects}. The sample used for our analysis is the SDSS DR5 Quasar
catalog (Schneider et al.~2007), as analyzed by Shen et al.~(2008)\footnote{We have verified that the results presented in this paper 
are globally unchanged if we use the new, updated catalogue by Shen et al. (2010) 
recently appeared in the literature. However, this update will be included in a forthcoming 
paper where we will analyze other consequences of the findings presented here.}. 
In order to work on a well-defined, high
quality sample, we further applied the following filters: redshift range between 0.01 and 0.8,
in order to have the [OIII] line fully inside the spectral range of optimal response; 
magnitude m$_i$$<$19.1, absolute magnitude M$_i$$<$22.1 (these two criteria define a
homogeneous, well selected subsample, consisting of more than half the total DR5 quasar
sample, Richards et al.~2006); signal-to-noise per pixel higher than 5, in order to have
high quality, reliable spectra. These criteria define a sample of $\sim$6,000 quasars,
which still allows a detailed analysis of the EW([OIII]) distribution.\\
The main issue relevant for our work is that the flux/luminosity limit has a strong influence
on the observed EW([OIII]) distribution. Qualitatively, we expect that for a given on-axis
flux/luminosity, there will be a maximum inclination angle, above which the object will
fall below one of the two limits. Therefore, the highest inclinations
are possible only in objects with on-axis flux/luminosity much above the sample limit. These objects
are obviously expected to be rare, due to the steepness of the quasar luminosity function.

In the following Section we quantitatively discuss these points, 
both analytically and through a Monte-Carlo simulation.
We will show that both methods clearly demonstrate that 
the expected slope of the distribution for high EWs is no longer dN/d(EW$_{O}$)$\propto$EW$_{O}^{-2}$,
as in the non-biased case described above, but dN/d(EW$_{O}$)$\propto$EW$_{O}^{-3.5}$.

\section{Data Analysis}
The expected shape of the EW([OIII]) distribution can be predicted through an analytic 
calculation, or through numerical simulations based on the observed data themselves.

{\bf Analytic calculation.} In the following, 
we use EW to refer to the [OIII] equivalent width, the subscript $O$ for observed
quantities, and $I$ for quantities, as would be measured in sources with a face-on disc. 
The observed luminosity L$_{O}$ is given by L$_{O}$=L$_{I}$$\times$$\cos \theta$, and 
the observed equivalent width EW$_{O}$=L([OIII])/L$_{O}$=EW$_{I}/\cos \theta$.


The differential number of objects dN with an intrinsic luminosity L$_I$, a "face on" equivalent
width EW$_I$, a disc inclination angle $\theta$, and a distance R, is:
\begin{equation}
dN = \Phi(L_I)dL_I \, g(EW_I)d(EW_I) \, d(cos \theta)\, R^2dR
\end{equation}
where $\Phi(L_I)$ is the intrinsic luminosity function and $g(EW_I)$ the distribution
of the intrinsic equivalent width.
We now want to transform this expression into a function of {\em observed} quantities,
through the variable change ($\cos \theta$, R)$\rightarrow$(EW$_O$, F), where
F is the observed flux. Given the Jacobian of the coordinate transformation, after straightforward algebra, we obtain:

\begin{equation}
dN \propto  \frac{L_I^{3/2}}{F^{5/2}EW_O} 
\left(\frac{EW_I}{EW_O}\right)^{5/2}\Phi(L_I)\,dEW_O\, dEW_I\, dL_I\, dF
\end{equation} 
If the luminosity function has the form $\Phi(L_I)\sim L_I^{-\beta}$, we can rearrange the above equation and obtain the observed EW distribution as:

\begin{eqnarray}
\frac{dN}{dEW_O} & \sim & \int_{L_{min}}^{L_{max}} L_I^{3/2-\beta}\,dL_I\int_{F_{min}}^{F_{max}}\frac{dF}{F^{5/2}}\times\nonumber\\
& & \times\int^{EW_O}_0 \frac{EW_I^{5/2}}
{EW_O^{7/2}}g(EW_I)\, dEW_I 
\end{eqnarray}
We have left $L_I$ as the integration variable instead of $L_O$, even if the
luminosity selection applies to the latter quantity. The reason is the
relatively flat slope $\beta$ below the break in the luminosity function:
the integral on $L_I$ is essentially dominated by the {\it upper} integration limit,
which is the {\it intrinsic} break luminosity and applies to $L_I$. Above the
break luminosity, the luminosity function is very steep; also, the integral on
EW$_I$ becomes independent of the upper limit EW$_O$ when the latter is much
larger than a typical EW$_I$. We thus obtain that  at the high EW$_O$ tail the observed distribution has
a power law shape 
\begin{equation}
\frac{dN}{dEW_O}\sim EW_O^{-3.5}
\end{equation}

{\bf Numerical estimate}. We now reproduce the same result through a simulation,
in order to have an independent check, and to test the effects of possible deviations 
of the intrinsic continuum luminosity function from the power law shape assumed above.
The simulation consists of the following steps:\\
1) We assume that our quasars have an
intrinsic Gaussian distribution of [OIII] equivalent width;
we further assume that the flux/luminosity distribution of our sample has the same shape 
it would have if one could use the intrinsic flux/luminosity instead of the observed
ones (i.e., at any observed flux/luminosity the sample is dominated by face-on
objects).\\
2) We extract a random object in our sample, and measure the observed continuum
luminosity L$_O$ at the line energy.\\
3) We select a random disc orientation, $\theta$, and derive a fictitious {\it
doubly-observed} continuum luminosity L$_F$=L$_O\times\cos \theta$,
and the analogous EW$_F$=EW$^*/\cos \theta$ (thus ignoring the actually measured value EW$_O$).\\
4) We apply an arbitrary rejection limit in terms of {\it doubly-observed} continuum flux and 
luminosity, analogous to the one applied to the original sample.\\
5) We repeat the procedure for a large (10$^6$) number of times, and 
analyze the distribution of EW$_F$.\\

The result of this exercise is a distribution with a high-EW tail dN/dEW$\sim$EW$^{\Gamma}$, 
with $\Gamma$=-3.50$\pm$0.01. 
We stress again that here we assume an intrinsic luminosity function with the same shape as the 
observed one. 
This is fully justified, since it is straightforward to show that starting from a population with a 
given intrinsic luminosity function, and assuming a random $\cos \theta$ correction, the shape of the 
observed luminosity function remains unchanged. This has been also checked with our {\it
doubly observed} data.

\subsection{A global fit to the EW distribution}

The main result of the above analysis is that, if we assume an isotropic emission of [OIII], proportional to the
intrinsic disc luminosity, and a random inclination of the disc with respect to the line of sight, the observed distribution of EW([OIII]) in
a flux-limited sample has a power-law tail towards high EWs with slope $\Gamma=$-3.5.

A simple look at Fig.~1 shows that such a tail is indeed present. This is
our main result, to be discussed in the next Section. Here we complete the analysis looking for
a global fit to the observed distribution, which is equivalent to finding the intrinsic distribution 
g(EW$_I$).

\begin{table}
\centerline{\begin{tabular}{lccc}
Parameter & Model 1 & Model 2 & Model 3 \\
\hline
EW$^*$(\AA)     & 7.9$\pm$0.4    & 7.1$\pm$0.3 & 8.0$\pm$0.3\\
$\sigma$(\AA)   & 8.5$\pm$1      & 4.7$\pm$0.2 & 4$\pm$0.3\\
EW$^*_2$(\AA)   & --             & --          & 17$\pm$1\\
$\sigma_2$(\AA) & --             & 9$\pm$1     & 11$\pm$0.8\\
$\alpha^a$      & --             & --          & 0.67$\pm$0.01\\
$\chi^2$/d.o.f. & 193/60         &126/59       & 46/57 \\
\end{tabular}}
\caption{Best fit parameters for the three models discussed in the text, consisting 
of the convolution of a power law with slope $\Gamma$=-3.5 with three different intrinsic distributions: a single symmetric Gaussian in Model~1,
a single asymmetric Gaussian in Model~3, two symmetric Gaussians in Model~3. $^a$: In model~3, the parameter $\alpha$ is the relative weight of the first Gaussian, i.e. $\alpha=N_1/(N_1+N_2)$, where N$_1$ and N$_2$ are the normalizations of the two Gaussians.}
\end{table}

We fitted the observed distribution by convolving g(EW$_I$) with the kernel describing the orientation 
effects, Equation~3, and assuming for g(EW$_I$) various modifications of Gaussian functions. In 
particular, we tried three different intrinsic distributions: a single symmetric Gaussian; an 
asymmetric Gaussian, with two different widths for values higher and lower than the average, respectively;
and two symmetric Gaussians. The first model has two free parameters in addition to a global normalization 
(the average EW$_I$ and the standard deviation $\sigma$); the second model has three parameters 
(EW$^*$ and the left and right standard deviations, $\sigma_L$ and $\sigma_R$); the third model has five 
parameters (EW$^*_{1}$, $\sigma_1$, EW$^*_{2}$, $\sigma_2$, and $\alpha$, the relative weight of the two 
distributions). In fitting the observed distribution, we used a $\chi^2$ minimization technique, assuming an error equal to the square root of the number counts. The width of each bin of the distribution is 1~\AA, smaller by a factor $\sim$2-3 than the typical error on single EW measurements, but large enough to have more than 15 counts in most bins, so that the use of Gaussian statistics is justified. The few bins with fewer counts have been rebinned in order to have at least 15 counts in each bin of the fitted distribution.
The results of the fits clearly favour the two-Gaussian model (Table~1).
The main differences between the fits 
are in the EW range around the distribution maximum, where a 
double slope change (characteristic of the two-Gaussian fit) is needed to reproduce properly the observed distribution (Fig.~1).
Instead, the high-EW tail is always well fitted by a $\Gamma$=-3.5 slope, and does not depend on the 
details of the intrinsic distribution. This is clear from a visual inspection of Fig.~1, where it is 
shown that the intrinsic distribution has an exponential drop at EW$\sim$20-30~\AA, and the tail at 
higher EW is entirely due to projection effects. This effect is even stronger for the single-peaked
models, which have a narrower intrinsic distribution (Table~1).  
Summarizing, the results of the global fit demonstrate that: (1) an intrinsic distribution of 
EW([OIII]), declining exponentially at high EWs, cannot reproduce the observed distribution, due to a prominent 
high-EW power law tail; 
(2) the tail of the distribution is completely 
explained by disc projection effects, independently of the detailed shape of the intrinsic 
distribution; (3) statistically, an intrinsic distribution made up of two populations different
in terms of [OIII] emission seems favoured with respect to single-peaked distributions.

The size of the sample (6,029 objects) allows an analysis of possible trends in the intrinsic distribution with
other physical parameters, in particular, the continuum luminosity. 
We divided the sample in three parts of equal number of objects, according to their oberved continuum luminosity,
and repeated the complete analysis of the EW([OIII]) distribution. In each case we obtained results fully compatible with those from the global fit, with each individual parameter consistent with the values in Table~1 at a 90\% confidence level. We conclude that there is no evidence of a dependence of the results of our analysis on the observed luminosity of the sample. 

\subsection{The distribution of broad lines EWs}

We repeated the analysis for the distributions of the equivalent widths of the main broad emission lines available in the 
SDSS DR5 sample, i.e. H$\beta$, Mg~II, and C~IV.
{The choice of these three lines is based on the availability of high-quality fits for a large sample of quasars (Shen et al.~2008). Moreover, they cover a large range in redshift (from z=0.01 to z=4.5) and luminosity, and a span a large range of ionization (from 7.6~eV for Mg~II to 13.6~eV for H$\beta$, to 48~eV for C~IV), possibly probing different zones of the broad line region. 
} 
The samples selection criteria are the same as for the [OIII] sample, with the only difference of the redshift range for the Mg~II 
(0.45$<$z$<$2.2) and the C~IV (1.7$<$z$<$4.5) line. 
The three samples consist of 6,029 (H$\beta$, the same as [OIII]), 19020 (Mg~II) and 4468 (C~IV) objects.
The main results of the global fits (Fig.~2) are the following.

- In all cases, an intrinsic distribution consisting of two Gaussians provide an excellent fit ($\chi^2$/d.o.f.$\sim$1), while single Gaussian distributions, either symmetric or asymmetric, are not statistically acceptable.\\
- For all the three EW distributions, the intrinsic distributions fall short of fitting the high-EW 
tail.
However, a power law with slope $\Gamma=-3.5$, as expected from disc projection effects
and a fully isotropic line emission, does not reproduce the
data either. A good fit can be obtained either with a steeper slope (Tab.~2) 
or, alternatively assuming a fixed $\Gamma$=-3.5, and a maximum inclination angle (cos $\theta \sim$ 0.2$\pm$0.03)
above which the disc is no longer visible (as one would expect 
if a thick torus is co-axial with the disc).

The above results clearly show that, regardless of the details of the {\em intrinsic} EW distributions, the {\em observed} 
EW distributions of the 
broad emission lines do not appear strongly affected by differential projection effects between 
lines and continuum. 

\section{Discussion}

We have shown that the distribution of EW([OIII])
in SDSS quasars is a strong, direct observational proof of the disc-like structure of the continuum emission 
source.
Furthermore, we have shown that a similar result does not hold for the main
broad emission lines (H$\beta$, Mg~II, C~IV). 
In the following we discuss these findings in more detail.

\subsection{The distribution of EW([OIII])}
The results obtained on the distribution of EW([OIII]) are interesting under several respects.
The central point stressed above is the direct observational evidence of disc inclination effects. 
In addition to this main point, we can gain
important insights on the structure of inner region of quasars.

{\bf Absence of a torus aligned with the disc.} One somewhat surprising result is the absence 
of a maximum angle of disc inclination in our quasar sample. If we assume an optically thick torus, 
co-aligned with the accretion disc, and not affecting the narrow-line region (i.e. the standard 
assumptions of the AGN unified model), we would expect a maximum inclination angle, above which 
the disc become invisible. In terms of the observed distribution, this would produce a sudden decline 
above some EW$_{\rm MAX}$, depending on the average torus opening angle (cos $\theta_{\rm MAX} 
\sim$EW$^*$/EW$_{ \rm MAX}$). Instead, no deviation from $\Gamma=-3.5$ is observed, which implies 
an upper limit on the maximum inclination angle cos $\theta_{\rm MAX}$$<$0.06.
So, either (1) the torus is randomly aligned with respect to the 
disc, or (2) the torus covering factor is extremely small in quasars.

\begin{table}
\centerline{\begin{tabular}{lccc}
Parameter & H$\beta$ & Mg~II  & C~IV \\
\hline
EW$_1^*$(\AA)     & 31$\pm$3      & 23.1$\pm$0.3   & 30$\pm$1     \\
$\sigma_1$(\AA)   & 11$\pm$1      & 5.0$\pm$0.2    & 8$\pm$1      \\
EW$_2^*$(\AA)     & 58$\pm$2      & 29$\pm$1       & 36$\pm$3     \\
$\sigma_2$(\AA)   & 15$\pm$1      & 8.4$\pm$0.4    & 16$\pm$3     \\
$\alpha$          & 0.20$\pm$0.07 & 0.63$\pm$0.07  & 0.62$\pm$0.17 \\
$\Gamma$          & -8.0$\pm$0.5  & -6.2$\pm$0.2   & -5.0$\pm$0.4  \\
$\chi^2$/d.o.f. & 89/97       & 78/67      & 61/72 \\
\end{tabular}}
\caption{Best fit parameters, for the convolution of a power law with a free slope 
$\Gamma$ and a two-Gaussian intrinsic distribution, for the observed 
EW distributions of the broad H$\beta$, Mg~II 
and C~IV lines. The parameter $\alpha$ is defined as $\alpha$=N$_1$/(N$_1$+N$_2$),
where N$_1$ and N$_2$ are the normalizations of the two Gaussians.
}
\end{table}

{\bf Effects of "limb darkening"}. We can predict the modifications to the observed EW([OIII]) due 
to possible limb darkening effects. If we parametrize the limb darkening with a parameter 
$\gamma$, so that the observed disc luminosity scales with the inclination angle as 
L($\theta$)$\sim$($\cos \theta)^{1+\gamma}$, we obtain from the calculation in Section~3 that the 
slope of the high-EW tail should change from $\Gamma=-3.5$ to $\Gamma=-3.5+\gamma$/(1+$\gamma$).
This deviation is not observed, putting an upper limit of $\gamma$$<$0.1 at a 90\% confidence limit.

{\bf Intrinsic distribution of EW([OIII]).}  The deconvolution of the projection effects and the 
determination of the intrinsic distribution of EW([OIII]) provides for the first time a 
quantitative estimate of the goodness of [OIII] as a proxy of the total AGN emission.

If we refer to the best fit with a single distribution (Model~2 in Table~1), the standard deviation 
is of the order of 50\%\ EW$^*$. Such a relatively large spread is expected, given the many 
parameters involved in the [OIII]-total luminosity relation. In particular, two aspects are relevant: 
(1) the covering factor and optical depth of the narrow-line clouds, and (2) the spread in the 
intrinsic quasar spectral energy distribution (Elvis et al.~1994). In principle, a dependence on 
the intrinsic SED can be assessed in our sample: since the
continuum radiation ionizing the [OIII] is in the UV range, the ratio between the [OIII] flux 
and the continuum at 5,000~\AA\ should be positively correlated with the UV/optical ratio. We 
investigated this possibility by plotting EW([OIII]) versus the ($u$-$g$) 
and ($u$-$r$) SDSS colours, and found no correlation. This suggests that the 
first point above (the size and composition of the narrow-line clouds) dominates the 
intrinsic spread of EW([OIII]).
\begin{figure}
\includegraphics[width=8.5cm,angle=0]{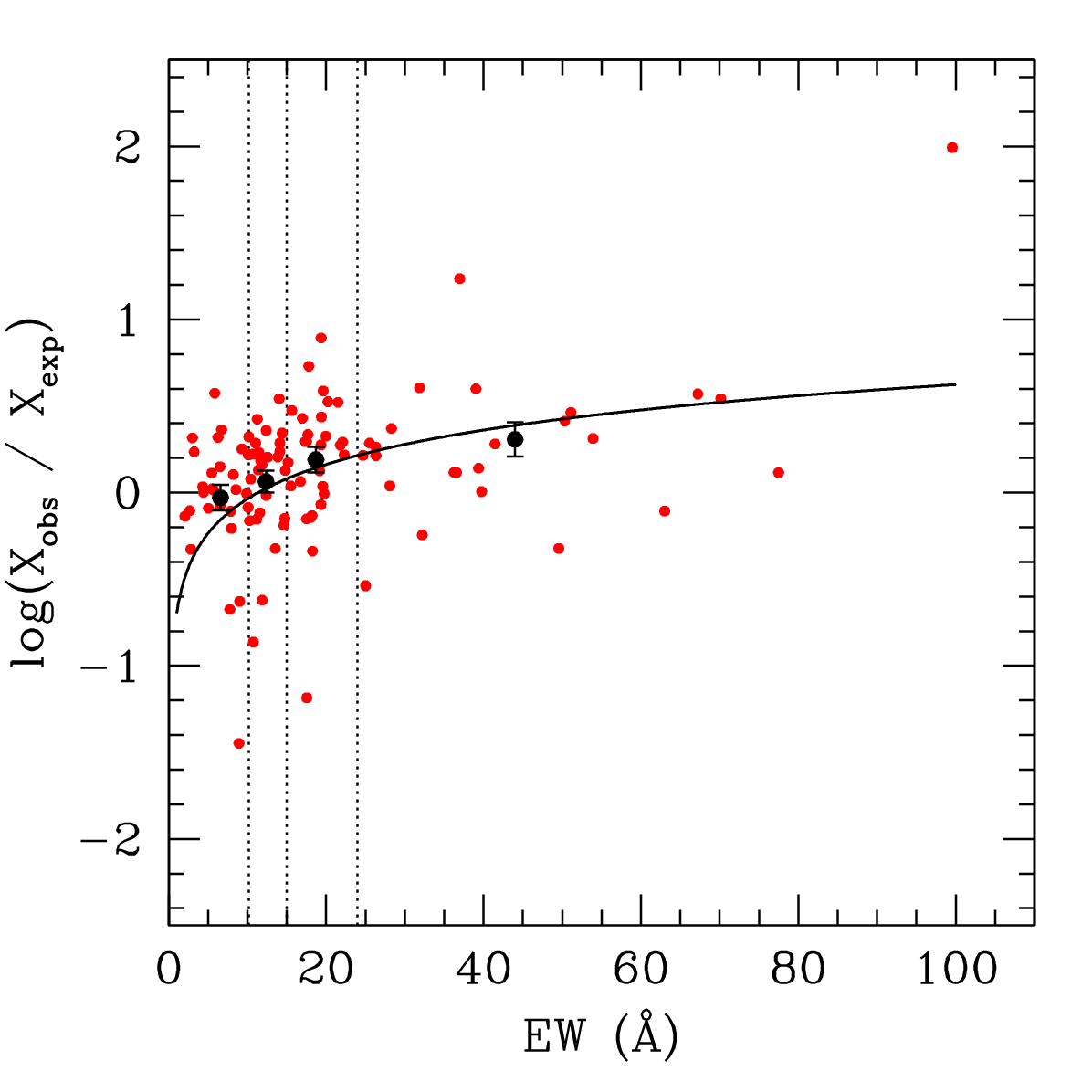}
\caption{Ratio between the observed 2-10~keV flux and the predicted value based 
on the $\alpha_{OX}$-L$_{UV}$ relation, versus EW([OIII]) for a sample of SDSS quasars with 
serendipitous XMM-Newton observations. Red points represent the average in four intervals (delimited by the  dotted lines) containing
the same number of objects. The continuous line shows the relation expected from projection effects.
}
\label{ttfit3}
\end{figure}

However, our best fit suggests the presence of two    
different populations in terms of [OIII] emission. 
If the two-population interpretation is correct, the calibration of 
the [OIII]  luminosity as an indicator of the total luminosity is more complex, and 
subject to higher uncertainties. Moreover, in this case other differences between the two 
populations should be found in a more complete investigation of the spectral properties of this 
sample. This is obviously an important issue, which will be treated in detail 
in a forthcoming work. Here we only note two straightforward points: (1) the exact nature 
of the intrinsic distribution does not affect the main result of the power-law shape of the high-EW 
tail { (see Section~4.2 for further details on this issue); }
(2) since there is no indication of a second population in the EW distributions of broad 
lines, the possible dichotomy observed in the [OIII] EW distribution should be due to differences
in the narrow-line region properties, rather than the disc properties.

{\bf Effects on the relation between optical and X-ray emission.} The hard X-ray emission of 
quasars is itself believed to be isotropic. Therefore, it 
can be used as an indicator of the bolometric luminosity,
 and is well known to 
correlate with the [OIII] emission (Heckman et al.~2005). 
The relation between bolometric and X-ray luminosity is 
complicated by the observed dependence of the optical/UV to X-ray ratio on the UV luminosity. 
The disc inclination obviously affects these correlations, adding a spread due to the projection 
effect on the UV continuum. In particular, objects which are seen nearly edge-on should have a 
higher (on average) X-ray to UV ratio. 
We note that overall we do not expect this to be a large contribution to the observed dispersion 
in the X-ray to UV ratio. The latter (Young et al.~2010) is of the order of a factor $\sim$3.
Considering that the observed cos $\theta$ distribution in our sample (and in any flux-limited sample, 
as shown in our calculation in Section~3) is N(cos $\theta$)$\sim(\cos \theta)^{1.5}$, 
we expect 
an average value of $<$cos $\theta$$>$=5/8, and a dispersion (defined as the interval containing 
68\% of the objects) of about 0.15, i.e. a factor of only $\sim 0.3 << 3$.
Nevertheless, in a large enough sample, a statistical correlation between the inclination angle 
and the observed X-ray to UV ratio should be measurable.

In order to check this, we performed the following test:\\
- We considered a subsample of the original 6,029 quasars, consisting of those (107) having
hard X-ray measurements. The X-ray data have been obtained from the Young et al.~(2009) catalog, 
made of all the SDSS DR5 quasars with serendipitous {\em XMM-Newton} observations.  \\
- We estimated for each object the expected 2-10~keV flux, based on the
$\alpha_{OX}$-UV luminosity correlation (Young et al.~2009). \\
- We plotted the ratio between the expected and measured X-ray fluxes, R=F$_{OBS}$/F$_{CALC}$, 
versus the measured EW([OIII]) for each object (Fig.~4).\\
- We calculated the expected R-EW relation, based on a N(cos $\theta$)$\sim$(cos $\theta$)$^{1.5}$ 
distribution of disc inclinations, and compared it with the observed values.\\

As shown in Fig.~4, even if the dispersion is large (see above), the bin-averaged R 
values follow the expected trend, starting below 1 at low EWs (i.e., more face-on than the average)
and growing above 1 at high EWs (ie, more edge-on). 

\begin{figure*}
\includegraphics[width=18.0cm,angle=0]{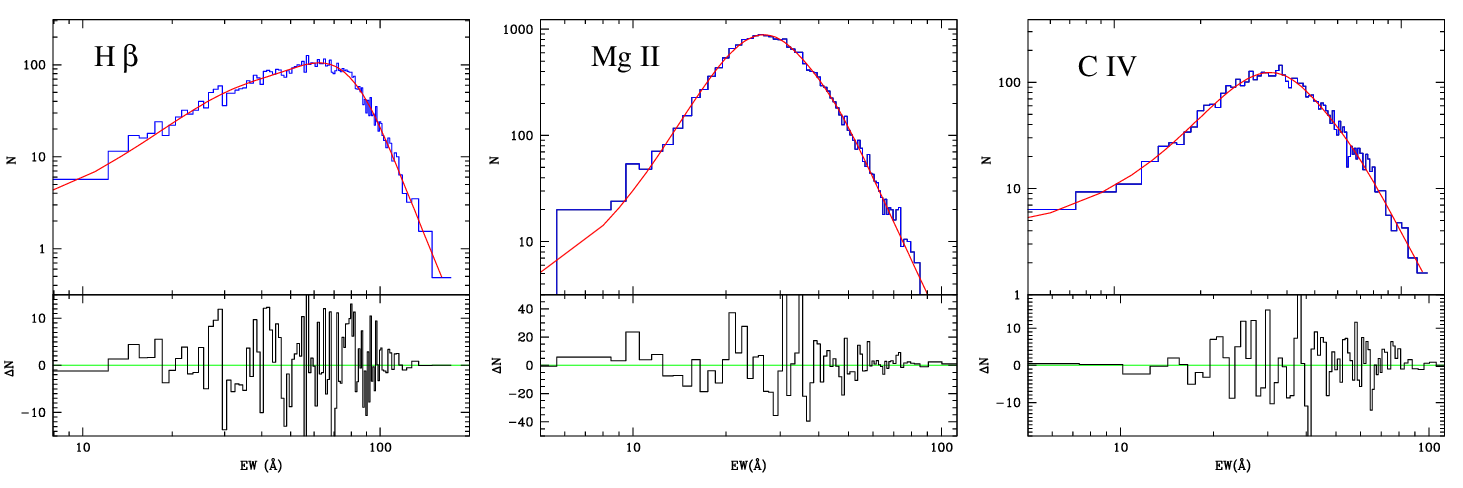}
\caption{Distributions of EW for three broad emission lines: H$\beta$ (left), 
Mg~II (center), and C~IV (right), with best fit models as described in the text and in Table~2.}
\label{ttfit4}
\end{figure*}

The result shown in Fig.~4 offers additional evidence in favour of our scheme. Since the presence of 
disc orientation effects were already proven by the analysis of the distribution of EW([OIII]), we can 
regard the trend in Fig.~4 as a suggestion of the hard X-ray emission being more isotropic than
the optical continuum.

\subsection{Uniqueness of orientation-based interpretation}

The model we have discussed so far provides a reasonable
interpretation of the data, but it is certainly not unique.
This notwithstanding, we will argue that it is the most
economic one, requiring a minimum number of ad hoc assumptions.

One possible alternative is to attribute the observed
distribution of EW([OIII]) not to inclination effects, but
to the intrinsic variance within the sample. A minimal
variance is already included in the model (the green line
in Fig.~1), but one could assume a larger one, so large
as to affect the slope of the high EW tail of the distribution. However,
any intrinsic distribution with a fall off steeper than
a $\Gamma$=-3.5 power law will leave unchanged the tail due
to inclination effects (provided that the sample extends to high 
enough EW); on the other hand, an intrinsic distribution 
with a fall off flatter than $\Gamma$=-3.5 will obviously produce
a similarly flat tail. In conclusion, the intrinsic hypothesis
would work only if (by chance) the intrinsic tail had exactly
the same slope as predicted by the inclination hypothesis.

Another alternative would attribute the dimming of the
observed continuum (and the enhancement of the observed
EW) not to inclination, but to absorption or blocking
mechanisms affecting preferentially the continuum with
respect to the [OIII]. The absorption/reddening hypothesis has 
already been advocated in connection with some "exceptional
[OIII] emitters" (e.g. Ludwig et al.~2010); we have checked this possibility
in our sample by looking for a correlation between EW
and continuum colors, but have found no evidence for it.
Any ad hoc escape would require a special kind of dust,
a grey absorber with a frequency-independent opacity.
As for the blocking variant, this has already been discussed
in connection with a possible circumnuclear torus (see
Section 4.1 above). In the simplest possible scheme (i.e.
torus opening identical in all sources, independent of 
redshift and/or luminosity) we have already seen that the 
prediction is a sharp drop in the EW distribution, and
have deduced from its non-observation a lower limit to
the opening angle. If one adopts a wide distribution of
opening angles, and assumes for the sake of simplicity that
the cumulative effect can be described as an "equivalent
limb darkening", then again we have found a stringent 
upper limit to this effect (see Section~4.1) if it were on top
of the inclination; if it were in lieu of the inclination,
the coincidence with the theoretical prediction ($\Gamma$=-3.5) 
would again remain unexplained.

A final option could be to relax the assumption of isotropy
we made for the [OIII] emission. This may happen if 
the
[OIII] emitting clouds extend so close to the exciting nucleus
as to be partly affected by the circumnuclear torus. If this
were indeed the case, the EW at high inclination would be
less than expected in our model, and the distribution tail
would become steeper (see Section~4.3 below, where a similar
scheme is discussed in connection with the EW distribution
of some broad lines). One could recover the observed EW([OIII])
distribution by assuming non-isotropy for both line and 
continuum. Once more, we describe the angular dependence
with a power law, and call $\alpha$ and $\beta$ the 
continuum and line exponent, respectively (F$_{OBS}$([OIII])$\sim$(cos$\theta)^\beta$,
F$_{OBS}$(CONT)$\sim$(cos$\theta)^\alpha$). 

One finds:
\begin{equation}
\frac{dN}{dEW}\sim EW^{-\frac{\frac{5}{2}\alpha+1-\beta}{\alpha-\beta}}
\end{equation}
{
The observations require $\Gamma$=-3.5$\pm$0.1 
which to first order implies
\begin{equation}
\alpha = 1 +\frac{5}{2}\beta \pm 0.1\times(1+\frac{3}{2}\beta)
\end{equation}
Only a narrow strip of values in the ($\alpha$, $\beta$) plane
are compatible with the data; of these, only our preferred
choice $\alpha$=1, $\beta$=0 has a clearcut physical justification.
 
It is of course possible that the actual values of $\alpha$ and $\beta$ 
differ slightly from the ``perfect disk'' ($\alpha$=1) and ``perfect
[O III] isotropy'' ($\beta$=0) scenario. 


It is also {\em formally} possible to assume for $\alpha$ and $\beta$
completely different values, as long as they satisfy Eq.~7.
However, such an ad-hoc assumption would not correspond to any
physically plausible configuration.

} 

\subsection{The EW distribution of broad emission lines}

The distributions of Fig.~3 do not show marked effects of different isotropy degrees
between broad lines and continuum, at variance with the
distribution of EW([OIII]). This suggests (1) a flattened structure of the broad line region, 
in order to have
(almost) the same projection effects in both the continuum and line emission, which would cancel out in 
the EW, and (2) a high optical depth of the emission lines.

We can further speculate on the differences  among the distributions of the three broad lines, with the 
EW(H$\beta$) one having the steepest high-EW slope.
This may be due to a somewhat rounder geometry of the line-emitting 
region (note however that Mg~II and C~IV have very different ionization levels, and in a
stratified medium should bracket the location of H$\beta$); or to a dependence of the emitting region 
geometry on the source redshift/luminosity: the redshift (and, in a flux-limited sample, the 
luminosity) is systematically different for the three broad-line samples considered here. In any 
case, the (second-order) differences among the EW distributions of the broad lines should not
mask the sharp difference between all of them and EW([OIII]).

{ The evidence for a flattened broad line region 
confirms early suggestions (Netzer 1987; Collin-Souffrin \& Dumont 1990; Wills \& Brotherton 1995; Wanders et al. 1995; Goad \& Wanders 1996) and more recent results, all based 
on different methods but pointing towards the same scenario:

- Spectropolarimetric observations  of Seyfert 1 galaxies and quasars show distinctive features across broad emission lines which can only 
be understood in terms of a rotating line-emitting disc (the BLR) surrounded by a coplanar scattering region (Smith et al. 2004, 2005).
In same cases, asymmetries in polarization spectra indicates that rotating winds are launched from these disks (Young et al. 2007).\\
- Maiolino et al. (2001) note an apparent paradox between the expected covering factor of BLR clouds ($\sim 30\%$) and the fact that the Ly-edge in absorption is never observed in quasar spectra. This paradox can be solved if the BLR is a disk and dusty gas in the outer parts, on the same plane, blocks observations along the lines of sight passing through the BLR clouds.\\
- McLure \& Dunlop~(2002) analyzed a sample of AGNs with black hole mass estimates from both reverberation mapping and stellar velocity dispersion, and showed that assuming a flattened distribution of the H$\beta$ emitting region provides a better match between the two estimates than a spherical shape.\\
- Jarvis \& McLure~(2006) found in a sample of radio-loud quasars a strong correlation between broad line widths and radio spectral index (considered an orientation indicator), suggesting a flattened shape of the broad line region.\\
- Down et al.~(2010) performed a detailed analysis of H$\alpha$ profiles in a sample of radio-loud, high-z quasars, and found that the complex observed profiles require that the line is emitted, at least in part, by a disk-like region.
}

The interpretation of the observed EW distributions of broad lines as due to a flattened emission region can be tested comparing the widths of the broad lines with EW([OIII]): being EW([OIII]) an indicator of the disc inclination, we expect on average a larger physical width of the broad lines in objects with higher EW([OIII]) (i.e. seen more edge-on).
This check can be done easily for the H$\beta$line width, which is available for the same sample providing the [OIII] measurements. The results are shown in Fig.~4. As for the case of the average X-ray emission shown in Fig.~2, the intrinsic dispersion of the W(H$\beta$)-EW([OIII]) relation is quite large, and visually hides any correlation; nevertheless a positive correlation { becomes apparent if we plot the averages of W(H$\beta$) for large enough EW([OIII]) intervals (Fig.~5). Quantitatively, a linear fit to the points provide a line correlation coefficient r=0.79, with a probability of null correlation P$<$10$^{-6}$.} 
This correlation further confirms the disc-like shape of the broad line emission region.  

\begin{figure}
\includegraphics[width=8.7cm,angle=0]{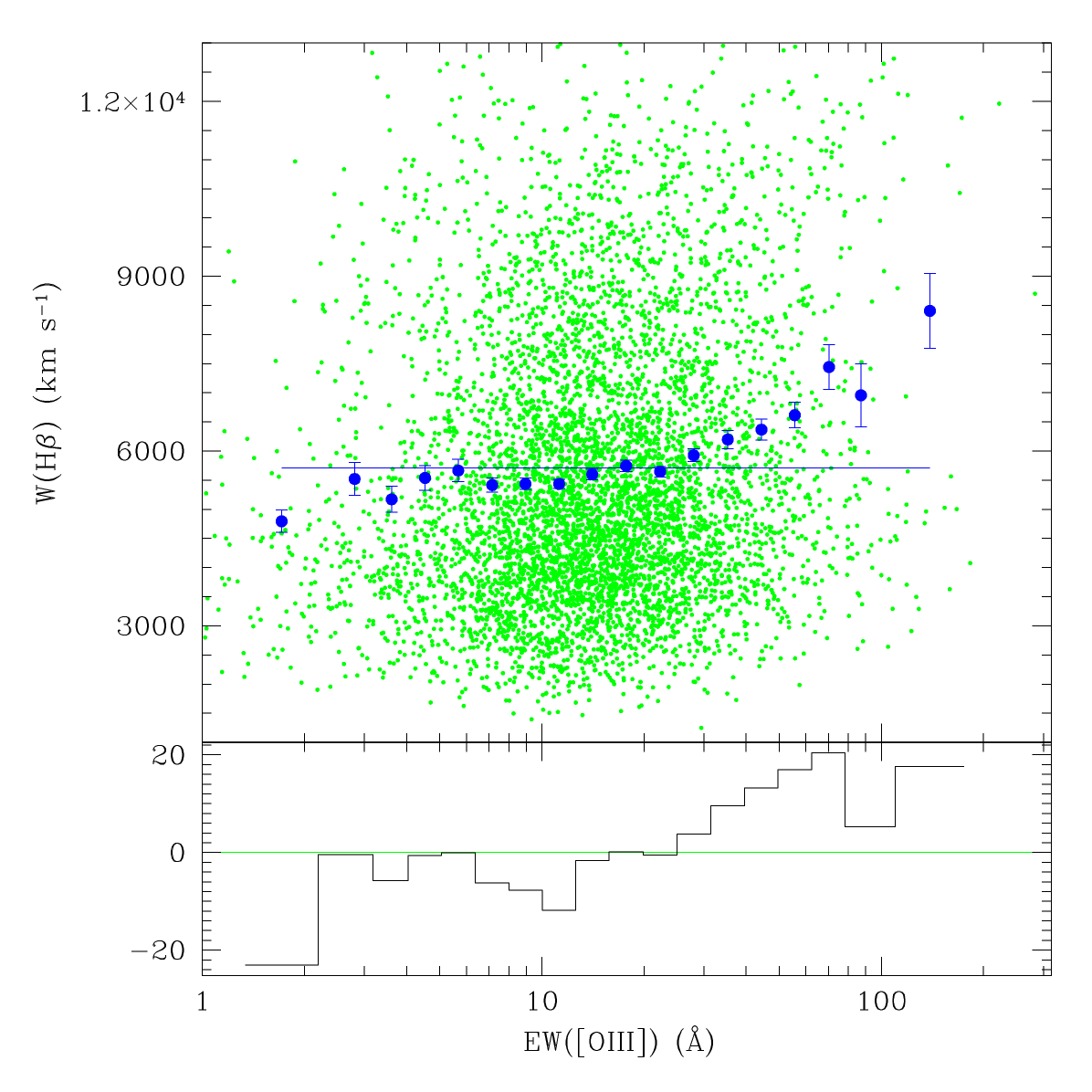}
\caption{H$\beta$ line width, W(H$\beta$), versus EW([OIII]) for our sample of 6,029 quasars. The small green points represent each object in the sample, while the large, blue points represent the average W(H$\beta$) in each logarithmic bin of EW([OIII]), with a width $\Delta(\log(EW))$=0.1. Lower panel: residuals with respect to the average of the whole sample (the horizontal line in the upper panel).
}
\label{ttfit4}
\end{figure}

\section{Conclusions}

We have presented an analysis of the EW distributions of the [OIII]~$\lambda$5007~\AA, H$\beta$, 
Mg~II~$\lambda$2800~\AA\ and C~IV~$\lambda$1549~\AA\  lines in flux-limited subsamples of
the SDSS DR5 quasar catalog. The main results are the following:\\
1) The distribution of EW([OIII]) exhibits a high-EW tail perfectly consistent with a model 
where the [OIII] emission is isotropic, and the continuum emission is due to a randomly oriented 
optically thick, geometrically thin disc. \\
2) The distribution of EW([OIII]) is not compatible with the presence of a torus co-aligned with 
the disc and covering more than a few degrees. \\
3) The deviations of the observed hard X-ray fluxes with respect to the average X-ray to UV 
correlation follow a trend with respect to EW([OIII]), suggesting that the X-ray emission is 
more isotropic than the optical continuum.\\ 
4) The EW distributions of the broad lines suggest that the broad line region has a flattened 
geometry, closer to that of the optical continuum emitting disc than to that of  the [OIII] emitting region. 
 
\section*{Acknowledgements}
We are grateful to the referee for his/her careful reading and
constructive comments.
This work has been partly
supported by NASA grants 
NNX10AF50G.


\end{document}